\documentclass[12pt]{article}
\usepackage{epsf}
\title{ {\bf
The effect of the Gaussian profile of the new Higgs doublet on the
radiative lepton flavor violating decay}}
\author{\vspace{1cm}\\
        {\bf E. O. Iltan}
        \thanks{E-mail address:
        eiltan@newton.physics.metu.edu.tr}
 \\
        Physics Department, Middle East Technical University \\
        Ankara, Turkey\\}

\date{}

\begin{document}
\setlength{\baselineskip}{24pt}
\maketitle
\setlength{\baselineskip}{7mm}
\begin{abstract}
We study the branching ratios of the lepton flavor violating
processes $\mu\rightarrow e\gamma$, $\tau\rightarrow e\gamma$  and
$\tau\rightarrow \mu\gamma$ by considering that the new Higgs
scalars localize with Gaussian profile in the extra dimension. We
see that the BRs of the LFV decays $\mu\rightarrow e \gamma$,
$\tau\rightarrow e \gamma$ and $\tau\rightarrow \mu\gamma$ are at
the order of the magnitude of $10^{-12}$, $10^{-16}$ and
$10^{-12}$ in the considered range of the free parameters. These
numerical values are slightly suppressed in the case that the
localization points of new Higgs scalars are different than
origin.
\end{abstract}
\thispagestyle{empty}
\newpage
\setcounter{page}{1}
\section{Introduction}
Lepton flavor violating (LFV) interactions reach great interest
since they are rich from the theoretical point of view. In the
standard model (SM), these decays are allowed by introducing the
neutrino mixing with non zero neutrino masses. However, their
branching ratios (BRs) are much below the experimental limits due
to the smallness of neutrino masses. Therefore, they are sensitive
the physics beyond the SM. Since, theoretically, the loop effects
are necessary for the existence of LFV decays, it would be
possible to predict the free parameters of the underlying theory
if one studies the measurable quantities of them.

Among the LFV processes, the radiative LFV $l_i\rightarrow
l_j\gamma$ $(i\neq j; i,j=e,\mu, \tau)$ decays deserve to analyze
and there are various experimental and theoretical works in the
literature. The current limits for the (BRs) of $\mu\rightarrow
e\gamma$ and $\tau\rightarrow e\gamma$ decays are $1.2\times
10^{-11}$ \cite{Brooks} and $3.9\times 10^{-7}$ \cite{Hayasaka},
respectively. A new experiment at PSI has been described and aimed
to reach to a sensitivity of $BR\sim 10^{-14}$ for $\mu\rightarrow
e\gamma$ decay \cite{Nicolo} and at present the experiment
(PSI-R-99-05 Experiment) is still running in the MEG
\cite{Yamada}. For $\tau\rightarrow \mu\gamma$ decay an upper
limit of $BR=9.0\, (6.8)\, 10^{-8}$ at $90\%$ CL has been obtained
\cite{Roney} (\cite{Aubert}), which is an improvement almost by
one order of magnitude with respect to previous one. From the
theoretical point of view, there is  an extensive work on the
radiative LFV decays in the literature
\cite{Barbieri1}-\cite{Paradisi}. In \cite{Barbieri1} these decays
were analyzed in the supersymmetric models. \cite{Iltan1, Diaz,
IltanExtrDim, IltanLFVSplit, IltanLFVSplitFat} and  \cite{Chang}
were devoted to the radiative LFV decays in the framework of the
two Higgs doublet model (2HDM) and in a model independent way,
respectively. In another work \cite{Paradisi}, they are analyzed
in the framework of the 2HDM and in the supersymmetric model.

In this work, we study the LFV processes $\mu\rightarrow e\gamma$,
$\tau\rightarrow e\gamma$  and $\tau\rightarrow \mu\gamma$ in the
2HDM with the inclusion of a single extra dimension. In the 2HDM
the radiative LFV decays are induced by the internal new neutral
Higgs bosons $h^0$ and $A^0$ and the extension of the Higgs sector
results in enhancement in the BRs of these decays. In addition to
this, the inclusion of a single extra dimension causes to modify
the BRs. The extra dimension scenario is based on the string
theories as a possible solution to the hierarchy problem of the
SM. The effects of extra dimensions on various phenomena have been
studied in the literature \cite{Arkani}-\cite{Lam}. In the extra
dimension scenarios the procedure is to pass from higher
dimensions to the four dimensions by compactifying each extra
dimension to a circle $S^1$ with radius $R$, which is a typical
size of corresponding extra dimension. This compactification leads
to appear new particles, namely Kaluza-Klein (KK) modes in the
theory. If all the fields feel the extra dimensions, so called
universal extra dimensions (UED), the extra dimensional momentum,
therefore the KK number at each vertex, is conserved. If the extra
dimensions are accessible to some fields but not all in the
theory, they are called non-universal extra dimensions. In this
case, the KK number at each vertex is not conserved and tree level
interaction of KK modes with the ordinary particles can exist. If
the fermions are assumed to locate at different points in the
extra dimension with Gaussian profiles, the hierarchy of fermion
masses can be obtained from the overlaps of fermion wave functions
and such scenario is called the split fermion scenario
\cite{Mirabelli}-\cite{Grossman}.

Here, we consider that the new Higgs doublet is localized in the
extra dimension with a Gaussian profile, by an unknown mechanism,
however, the other particle zero modes have uniform profile in the
extra dimension. The Higgs localization in the extra dimension has
been considered in several works. The work \cite{IltanLFVSplitFat}
was devoted to the branching ratios of the radiative LFV decays in
the split fermion scenario, with the assumption that the new Higgs
doublet is restricted to the 4D brane or to a part of the bulk in
one and two extra dimensions, in the framework of the 2HDM. The
idea of the localization of the SM Higgs, using the localizer
field, has been studied in \cite{Surujon}. In
\cite{iltSplitHiggsLocal} the new Higgs scalars were localized in
the extra dimension and the SM Higgs was considered to have a
constant profile. The localization of new Higgs scalars depended
strongly on the strength of the small coupling of the localizer to
the new Higgs scalar. In the present analysis, we first consider
that the new Higgs scalars localize with Gaussian profiles around
origin in the extra dimension. Second, we assume that the
localization point is different than the origin but near to that.
We see that the BRs of the LFV decays $\mu\rightarrow e \gamma$,
$\tau\rightarrow e \gamma$ and $\tau\rightarrow \mu\gamma$ are at
the order of the magnitude of $10^{-12}$, $10^{-16}$ and
$10^{-12}$ in the given range of the free parameters. These
numerical values are slightly suppressed in the case that the
localization points of new Higgs scalars are different than
origin.

The paper is organized as follows: In Section 2, we present the
lepton-lepton-new Higgs scalar vertices and the BRs of the
radiative LFV decays with the assumption that the new Higgs
doublet is localized with a Gaussian profile in the extra
dimension in the 2HDM. Section 3 is devoted to discussion and our
conclusions.
\section{The effect Gaussian profile of new Higgs scalars
on the radiative LFV decays in the 2HDM }
The radiative LFV $l_i\rightarrow l_j\,\gamma$ decays are rare
decays in the sense that they exist at loop level in the SM. Since
the numerical values of the BRs of these decays are extremely
small in the framework of the SM, one goes the models beyond where
the particle spectrum is extended and the additional contributions
result in an enhancement in the numerical values of the physical
parameters. Due to the extended Higgs sector, the version of the
2HDM, permitting the existence of the FCNCs at tree level, is one
of the candidate to obtain relatively large BRs of the decays
under consideration. Furthermore, we take into account the effects
of the inclusion of a single spatial extra dimension which causes
to enhance the BRs due to the fact that the particle spectrum is
further extended after the compactification. Here, we consider the
effects of the additional Higgs sector with the assumption that
the new Higgs scalar zero modes are localized in the extra
dimension with Gaussian profiles by an unknown mechanism, on the
other hand, the zero modes of other particles have uniform profile
in the extra dimension. The Yukawa Lagrangian responsible for the
LFV interactions in a single extra dimension reads,
\begin{eqnarray}
{\cal{L}}_{Y}=
\xi^{E}_{5\,\,ij} \bar{l}_{i L} \phi_{2} E_{j R} + h.c. \,\,\, ,
\label{lagrangian}
\end{eqnarray}
where $L$ and $R$ denote chiral projections $L(R)=1/2(1\mp
\gamma_5)$, $\phi_{2}$ is the new scalar doublet and $\xi^{E}_{5\,
ij}$ are the FV complex Yukawa couplings in five dimensions, where
$i,j$ are family indices of leptons, $\phi_{i}$ for $i=1,2$, are
the two scalar doublets, $l_{i}$ and $E_{j}$ are lepton doublets
and singlets respectively. These fields are the functions of
$x^\mu$ and $y$, where $y$ is the coordinate represents the fifth
dimension.

We choose the Higgs doublets $\phi_{1}$ and $\phi_{2}$ as
\begin{eqnarray}
\phi_{1}=\frac{1}{\sqrt{2}}\left[\left(\begin{array}{c c}
0\\v+H^{0}\end{array}\right)\; + \left(\begin{array}{c c} \sqrt{2}
\chi^{+}\\ i \chi^{0}\end{array}\right) \right]\, ;
\phi_{2}=\frac{1}{\sqrt{2}}\left(\begin{array}{c c} \sqrt{2}
H^{+}\\ H_1+i H_2 \end{array}\right) \,\, , \label{choice}
\end{eqnarray}
and their vacuum expectation values read:
\begin{eqnarray}
<\phi_{1}>=\frac{1}{\sqrt{2}}\left(\begin{array}{c c}
0\\v\end{array}\right) \,  \, ; <\phi_{2}>=0 \,\, .
\label{choice2}
\end{eqnarray}
In this case, it is possible to collect the SM (new) particles in
the first (second) doublet and $H_1$ and $H_2$ becomes the mass
eigenstates $h^0$ and $A^0$, respectively since no mixing occurs
between two CP-even neutral bosons $H^0$ and $h^0$ at tree level.
%
%

The five dimensional lepton doublets and singlets have both
chiralities and the four dimensional Lagrangian is constructed by
expanding these  fields into their KK modes. Besides, the extra
dimension denoted by $y$ is compactified on an orbifold $S^1/Z_2$
with radius $R$. The KK decompositions of the lepton and the SM
Higg fields read
\begin{eqnarray}
\phi_{1}(x,y ) & = & {1 \over {\sqrt{2 \pi R}}} \left\{
\phi_{1}^{(0)}(x) + \sqrt{2}
\sum_{n=1}^{\infty}  \phi_{1}^{(n)}(x) \cos(ny/R)\right\}\, ,\nonumber\\
l_i (x,y )& = & {1 \over {\sqrt{2 \pi R}}} \left\{ l_{i
L}^{(0)}(x) + \sqrt{2} \sum_{n=1}^{\infty} \left[l_{i L}^{(n)}(x)
 \cos(ny/R) + l_{i R}^{(n)}(x) \sin(ny/R)\right]\right\}\, ,\nonumber\\
E_{i}(x,y )& = & {1 \over {\sqrt{2 \pi R}}} \left\{ E_{i
R}^{(0)}(x) + \sqrt{2} \sum_{n=1}^{\infty}  \left[E_{i R}^{(n)}(x)
\cos(ny/R) + E_{i L}^{(n)}(x) \sin(ny/R)\right]\right\} \,\, ,
\label{f0}
\end{eqnarray}
where $\phi_{1}^{(0)}(x)$, $l_{i L}^{(0)}(x)$ and $E_{i
R}^{(0)}(x)$ are the four dimensional Higgs doublet, lepton
doublets and lepton singlets respectively. Here, we assume that
the new Higgs scalars are localized in the extra dimension with
Gaussian profiles,
\begin{eqnarray}
S(x,y )=A e^{-\beta y^2} S(x) \label{phi2}\, ,
\end{eqnarray}
by an unknown mechanism\footnote{We consider the zero mode Higgs
scalars and we do not take into account the possible KK modes of
Higgs scalars since the mechanism for the localization is unknown
and we expect that the those contributions are small due to their
heavy masses.}. The normalization constant $A$ is
\begin{eqnarray}
A=\frac{(2\,
\beta)^{1/4}}{\pi^{1/4}\,\sqrt{Erf[\sqrt{2\,\beta}\,\pi\,R]}}\label{Norm}
\, ,
\end{eqnarray}
and the parameter $\beta=1/\sigma^2$ regulates the amount of
localization, where $\sigma$, $\sigma=\rho\,R$, is the Gaussian
width of $S(x,y)$ in the extra dimension. Here the function
$Erf[z]$ is the error function, which is defined as
\begin{eqnarray}
Erf[z]=\frac{2}{\sqrt{\pi}}\,\int_{0}^{z}\,e^{-t^2}\,dt \,\, .
\label{erffunc}
\end{eqnarray}
The coupling of the new Higgs doublet to the leptons brings
modified Yukawa interactions in four dimensions. To obtain the
lepton-lepton-Higgs interaction coupling in four dimensions we
need to integrate the combination $\bar{f}^{(0
(n))}_{iL\,(R)}(x,y)\,S(x,y)\, f^{(n (0))}_{j R\,(L)}(x,y)$,
appearing in the part of the Lagrangian (eq. (\ref{lagrangian})),
over the fifth dimension. Using the KK basis for lepton fields
(see eq. (\ref{f0})), we get
\begin{eqnarray}
\int_{-\pi R}^{\pi R}\, dy\,\, \bar{f}^{(0
(n))}_{iL\,(R)}(x,y)\,S(x,y)\, f^{(n (0))}_{jR\,(L)}(x,y)=V_n \,
\bar{f}^{(0(n))}_{iL\,(R)}(x) \,S(x)\,\,f^{(n (0))}_{j
R\,(L)}(x)\,\, , \label{intVij1}
\end{eqnarray}
where the factor $V_n$ reads
\begin{eqnarray}
V_n=A \, c_n \, , \label{Vij1even}
\end{eqnarray}
and the function $A$ is defined in eq. (\ref{Norm}). Here, the
fields $f^{(n (0))}_{iL}$, $f^{(n (0))}_{i R}$ are four
dimensional left and right handed zero (n) mode lepton fields. The
function $c_n$ in eq. (\ref{Vij1even}) is obtained as:
\begin{eqnarray}
c_ n=e^{-\frac{n^2}{4\,\beta\,R^2}}\, \frac{\Bigg( Erf[\frac{i\,
n+2\,\beta\,\pi\,R^2}{2\,\sqrt{\beta}\,R}]+Erf[\frac{-i\,
n+2\,\beta\,\pi\,R^2}{2\,\sqrt{\beta}\,R}]\Bigg)}
{4\,\sqrt{\beta\,\pi}\,R} \label{cevenodd} \, ,
\end{eqnarray}
Notice that the Yukawa couplings $\xi^{E}_{ij}$ in four dimensions
are
\begin{eqnarray}
\xi^{E}_{ij}= A \, \xi^{E}_{5\, ij}\, , \label{coupl4}
\end{eqnarray}
where $\xi^{E}_{5\, ij}$ are  the Yukawa couplings in five
dimensions (see eq. (\ref{lagrangian}))\footnote{In the following
we use the dimensionful coupling $\bar{\xi}^{E}_{N,ij}$ in four
dimensions, with the definition $\xi^{E}_{N,ij}=\sqrt{\frac{4\,
G_F}{\sqrt{2}}}\, \bar{\xi}^{E}_{N,ij}$ where N denotes the word
"neutral".}.

Now, we consider that the new Higgs scalars are localized in the
extra dimension at the point $y_H$, $y_H=\alpha\, R$ near to the
origin, namely,
\begin{eqnarray}
S(x,y)=A_H \,e^{-\beta (y-y_H)^2}\, S(x) \label{phi2H}\, ,
\end{eqnarray}
with the normalization constant
\begin{eqnarray}
A_H=\frac{2\,( \beta)^{1/4}}{(2
\pi)^{1/4}\,\sqrt{Erf[\sqrt{2\,\beta}\,(\pi\,R+y_H)]+
Erf[\sqrt{2\,\beta}\,(\pi\,R-y_H)]}} \label{NormH} \, .
\end{eqnarray}
After integrating the combination $\bar{f}^{(0
(n))}_{iL\,(R)}(x,y)\,S(x,y)\, f^{(n (0))}_{j R\,(L)}(x,y)$ over
extra dimension, the factor $V_n$ in eq. (\ref{intVij1}) reads
\begin{eqnarray}
V_n=A_H \, c_n \, , \label{Vij1evenH}
\end{eqnarray}
with the function $A_H$ in eq. (\ref{NormH}). The function $c_n$
in eq. (\ref{Vij1evenH}) is calculated as:
\begin{eqnarray}
c_ n=e^{-\frac{n^2}{4\,\beta\,R^2}}\, Cos[\frac{n\,y_H}{R}]\,
\frac{\Bigg( Erf[\frac{i\,
n+2\,\beta\,\pi\,R^2}{2\,\sqrt{\beta}\,R}]+Erf[\frac{-i\,
n+2\,\beta\,\pi\,R^2}{2\,\sqrt{\beta}\,R}]\Bigg)}
{4\,\sqrt{\beta\,\pi}\,R} \label{cevenoddH} \, .
\end{eqnarray}
Similar to the previous case, we define the Yukawa couplings in
four dimensions as
\begin{eqnarray}
\xi^{E}_{ij}= A_H \, \xi^{E}_{5\, ij}\, . \label{coupl4H}
\end{eqnarray}

\begin{figure}[htb]
\vskip -4.0truein \epsfxsize=4.8in\hskip 1.7truein
\leavevmode\epsffile{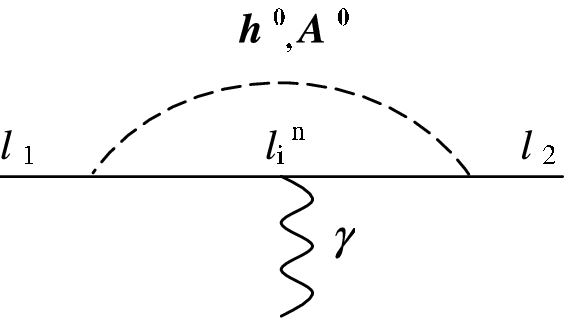} \vskip 10.0truein \caption[]{One
loop diagrams contribute to $l_1\rightarrow l_2 \gamma$ decay  due
to the zero mode neutral Higgs bosons $h^0$ and $A^0$ in the 2HDM,
for a single extra dimension. Here $l_i^n$ represents the internal
KK mode charged lepton and n=0,1, ...} \label{fig1}
\end{figure}

Now, we will present the decay widths of the LFV processes
$\mu\rightarrow e\gamma$, $\tau\rightarrow e\gamma$ and
$\tau\rightarrow \mu\gamma.$ These decays exist at least at one
loop level in the 2HDM and there appear the logarithmic
divergences in the calculations. These divergences can be
eliminated by using the on-shell renormalization
scheme\footnote{In this scheme, the self energy diagrams for
on-shell leptons vanish since they can be written as $
\sum(p)=(\hat{p}-m_{l_1})\bar{\sum}(p) (\hat{p}-m_{l_2})\, , $
however, the vertex diagrams (see Fig.\ref{fig1}) give non-zero
contribution. In this case, the divergences can be eliminated by
introducing a counter term $V^{C}_{\mu}$ with the relation
$V^{Ren}_{\mu}=V^{0}_{\mu}+V^{C}_{\mu} \, , $ where
$V^{Ren}_{\mu}$ ($V^{0}_{\mu}$) is the renormalized (bare) vertex
and by using the gauge invariance: $k^{\mu} V^{Ren}_{\mu}=0$.
Here, $k^\mu$ is the four momentum vector of the outgoing
photon.}. The decay width $\Gamma$ for the $l_i\rightarrow
l_j\gamma$ decay reads
\begin{eqnarray}
\Gamma (l_i\rightarrow l_j\gamma)=c_1(|A_1|^2+|A_2|^2)\,\, ,
\label{DWmuegam}
\end{eqnarray}
for $l_i\,(l_j)=\tau;\mu\,(\mu$ or $e; e)$. Here $c_1=\frac{G_F^2
\alpha_{em} m^3_{l_i}}{32 \pi^4}$, $A_1$ ($A_2$) is the left
(right) chiral amplitude and taking only $\tau$ lepton for the
internal line\footnote{We take into account only the internal
$\tau$-lepton contribution since, we respect the Sher scenerio
\cite{Sher}, results in the couplings $\bar{\xi}^{E}_{N, ij}$
($i,j=e,\mu$) are small compared to $\bar{\xi}^{E}_{N,\tau\, i}$
$(i=e,\mu,\tau)$, due to the possible proportionality of them to
the masses of leptons under consideration in the vertices.}, they
read
\begin{eqnarray}
A_1&=&Q_{\tau} \frac{1}{48\,m_{\tau}^2} \Bigg \{ 6\,m_\tau\,
\bar{\xi}^{E *}_{N,\tau f_2}\, \bar{\xi}^{E *}_{N,f_1\tau}\,\Bigg(
c_0^2\, \Big( F (v_{h^0})-F (v_{A^0})\Big) \nonumber \\ &+& 2\,
\sum_{n=1}^{\infty}\, c_n^2 \Big( F (v_{n, h^0})-F (v_{n,
A^0})\Big ) \Bigg ) + m_{f_1}\,\bar{\xi}^{E *}_{N,\tau f_2}\,
\bar{\xi}^{E}_{N,\tau f_1}\, \Bigg( c_0^2\,
\Big( G (v_{h^0})+G(v_{A^0}) \Big )\nonumber \\
&+& 2\,\sum_{n=1}^{\infty}\, c_n^2\, \Big( G (v_{n, h^0})+G (v_{n,
A^0})\Big)\Bigg) \Bigg \}
\nonumber \,\, , \\
A_2&=&Q_{\tau} \frac{1}{48\,m_{\tau}^2} \Bigg \{ 6\,m_\tau\,
\bar{\xi}^{E}_{N, f_2 \tau}\, \bar{\xi}^{E}_{N,\tau f_1}\,\Bigg(
c_0^2\,
\Big(F(v_{h^0})-F(v_{A^0})\Big)\nonumber \\
&+& 2\,\sum_{n=1}^{\infty}\,c_n^2\,\Big( F (v_{n, h^0})-F (v_{n,
A^0})\Big) \Bigg)+ m_{f_1}\,\bar{\xi}^{E}_{N,f_2\tau}\,
\bar{\xi}^{E *}_{N,f_1
\tau}\,\Bigg( c_0^2 \,\Big( G (v_{h^0})+G (v_{A^0})\Big) \nonumber \\
&+& 2\,\sum_{n=1}^{\infty}\,c_n^2 \, \Big(G (v_{n, h^0})+ G (v_{n,
A^0})\Big) \Bigg) \Bigg\}
 \,\, , \label{A1A22}
\end{eqnarray}
where $v_{n, S}=\frac{m^2_{\tau}+m_n^2}{m^2_{S}}$,
$m_n=\frac{n}{R}$ and $Q_{\tau}$ is the charge of $\tau$ lepton.
Here the vertex factor $c_n$  is defined in eq. (\ref{cevenodd})
(eq. (\ref{cevenoddH})) and the functions $F (w)$ and $G (w)$ are
\begin{eqnarray}
F (w)&=&\frac{w\,(3-4\,w+w^2+2\,ln\,w)}{(-1+w)^3} \, , \nonumber \\
G (w)&=&\frac{w\,(2+3\,w-6\,w^2+w^3+ 6\,w\,ln\,w)}{(-1+w)^4} \,\,
. \label{functions2}
\end{eqnarray}
%
\section{Discussion}
In the present work, we study the radiative LFV decays
$l_i\rightarrow l_j\gamma$ in the 2HDM with the addition of a
single spatial extra dimension. Here, we consider that new Higgs
scalars are localized in the extra dimension with Gaussian
profiles by an unknown mechanism, on the other hand, the other
particles have uniform zero mode profiles in the extra dimension
which is compactified on to orbifold $S_1/Z_2$. Since these decays
exist at least at one loop level, there appear free parameters
related to the model used in the theoretical values of the
physical quantities. The Yukawa couplings $\bar{\xi}^E_{N,ij}, \,
i,j=e, \mu, \tau$ are among those parameters. We consider that the
couplings $\bar{\xi}^{E}_{N,ij},\, i,j=e,\mu $ are smaller
compared to $\bar{\xi}^{E}_{N,\tau\, i}\, i=e,\mu,\tau$ since
latter ones contain heavy flavor. Furthermore, we assume that, in
four dimensions, the couplings $\bar{\xi}^{E}_{N,ij}$ are
symmetric with respect to the indices $i$ and $j$. For the Higgs
masses we take the numerical values $m_{h^0}=100\, GeV$,
$m_{A^0}=200\, GeV$. The compactification scale $1/R$ and the
Gaussian width $\sigma$ of new Higgs doublet are the additional
free parameters which are chosen not to contradict with the
experimental results. The direct limits from searching for KK
gauge bosons imply $1/R> 800\,\, GeV$, the precision electro weak
bounds on higher dimensional operators generated by KK exchange
place a far more stringent limit $1/R> 3.0\,\, TeV$ \cite{Rizzo}
and, from $B\rightarrow \phi \, K_S$, the lower bounds for the
scale $1/R$ have been obtained as $1/R > 1.0 \,\, TeV$, from
$B\rightarrow \psi \, K_S$ one got $1/R
> 500\,\, GeV$, and from the upper limit of the $BR$, $BR \, (B_s
\rightarrow \mu^+ \mu^-)< 2.6\,\times 10^{-6}$, the estimated
limit was $1/R > 800\,\, GeV$ \cite{Hewett}. Here, we take the
compactification scale $1/R$ in the range $200\, GeV\leq 1/R \leq
1000\, GeV$ and choose the Gaussian width at most $\rho=0.05$.

Our analysis is based on the Higgs localization width and the
compactification scale dependence of the BRs of the LFV decays.
First, we consider that the new Higgs is localized around the
origin in the extra dimension. Furthermore, we choose the
localization point is near to the origin and study its effect on
the BRs.

\begin{figure}[htb]
\vskip -3.0truein \centering \epsfxsize=6.8in
\leavevmode\epsffile{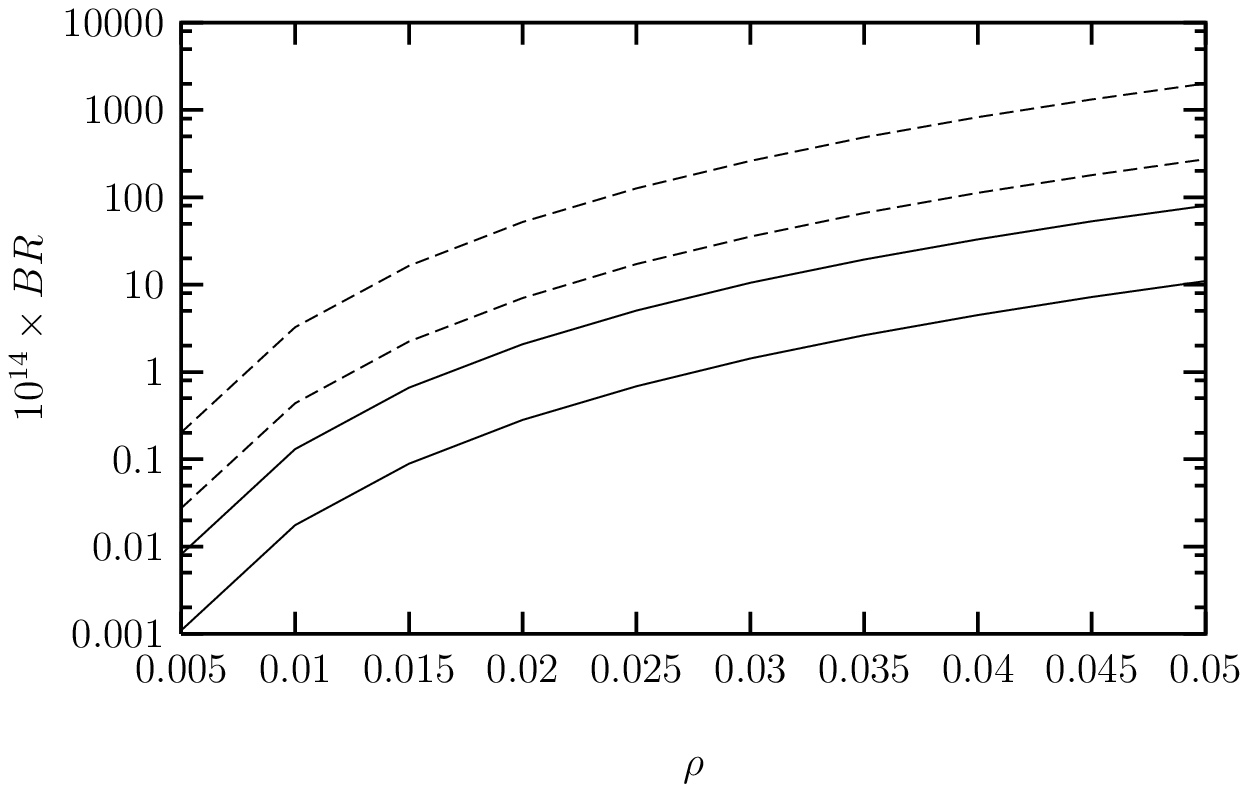} \vskip -3.0truein \caption[]{The
BR($\mu\rightarrow e \gamma$) with respect to $\rho$. Here the
lower-upper solid (dashed), line represents the BR for a single
extra dimension without-with lepton KK modes, the real couplings
$\bar{\xi}^{E}_{N,\tau \mu} =50\, GeV$, $\bar{\xi}^{E}_{N,\tau e}
=0.1\, GeV$ ($\bar{\xi}^{E}_{N,\tau e} =0.5\, GeV$) and
$R=0.005\,GeV^{-1}$. } \label{muegamro}
\end{figure}

Fig. \ref{muegamro} represents the BR($\mu\rightarrow e \gamma$)
with respect to $\rho$, for different values of the Yukawa
couplings $\bar{\xi}^{E}_{N,\tau \mu}$ and $\bar{\xi}^{E}_{N,\tau
e}$. Here the lower-upper solid (dashed) line represents the BR
for a single extra dimension without-with lepton KK modes, the
real couplings $\bar{\xi}^{E}_{N,\tau \mu} =50\, GeV$,
$\bar{\xi}^{E}_{N,\tau e} =0.1\, GeV$ ($\bar{\xi}^{E}_{N,\tau e}
=0.5\, GeV$) and $R=0.005\,GeV^{-1}$. It is observed that the BR
is strongly sensitive to the Gaussian width of the localized
neutral Higgs scalars and it increases with the increasing values
of the width. This enhancement is almost at the order of $10^4$ in
the range $0.005\leq \rho \leq 0.05$. The inclusion of lepton KK
modes brings additional enhancement at one order. The BR is at the
order of the magnitude of $10^{-14}$ ($10^{-12}$) for the
intermediate values of the localization parameter $\rho$ and the
coupling $\bar{\xi}^{E}_{N,\tau e} =0.1\, GeV$
($\bar{\xi}^{E}_{N,\tau e} =0.5\, GeV$). A new experiment at PSI
has been described and aimed to reach to a sensitivity of $BR\sim
10^{-14}$ and at present the experiment (PSI-R-99-05 Experiment)
is still running in the MEG \cite{Yamada}. The improvement of the
numerical result of the BR would make it possible to search the
effects of the extra dimensions and the possible localization of
new Higgs bosons in the extra dimension.

\begin{figure}[htb]
\vskip -3.0truein \centering \epsfxsize=6.8in
\leavevmode\epsffile{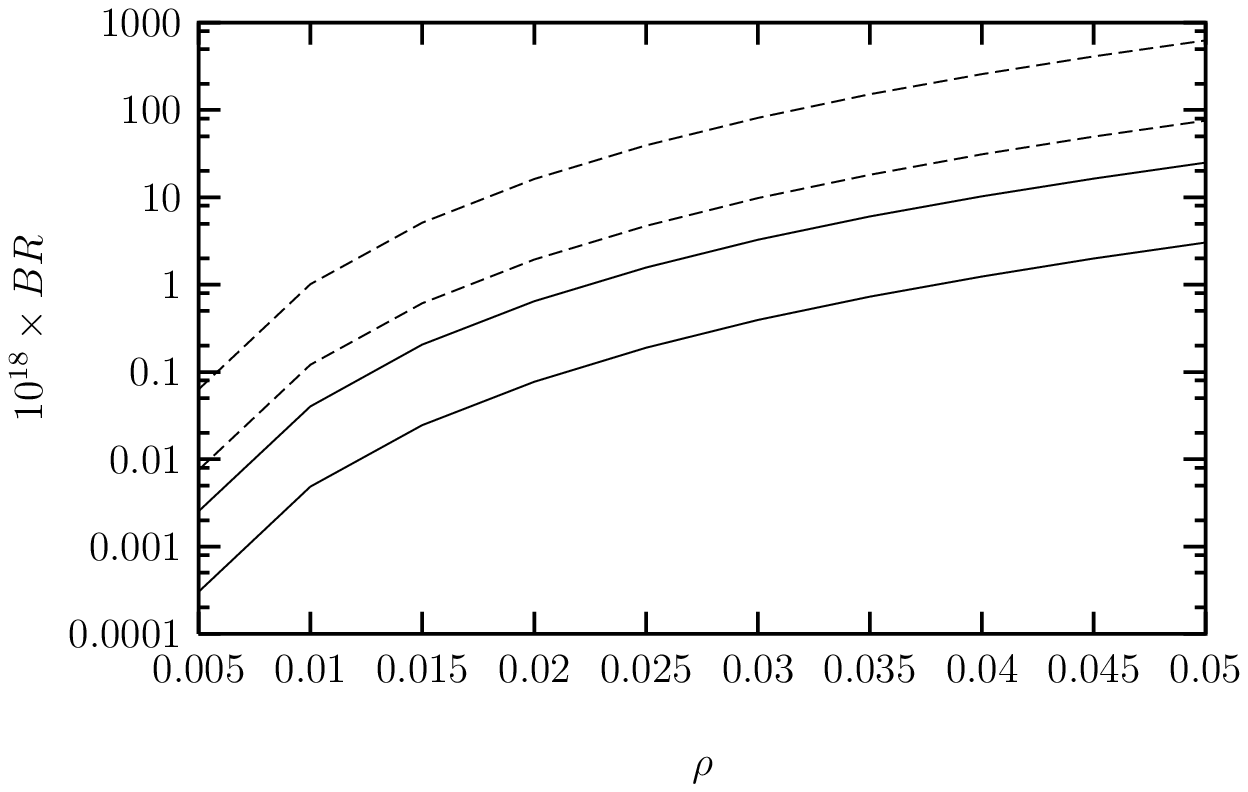} \vskip -3.0truein
\caption[]{The BR($\tau\rightarrow e \gamma$)  with respect to
$\rho$. Here the lower-upper solid (dashed), line represents the
BR for a single extra dimension without-with lepton KK modes, for
$R=0.005\,GeV^{-1}$, the real couplings $\bar{\xi}^{E}_{N,\tau
\tau} =100\, GeV$, $\bar{\xi}^{E}_{N,\tau e} =0.1\, GeV$
($\bar{\xi}^{E}_{N,\tau e} =0.5\, GeV$). } \label{tauegamro}
\end{figure}

\begin{figure}[htb]
\vskip -3.0truein \centering \epsfxsize=6.8in
\leavevmode\epsffile{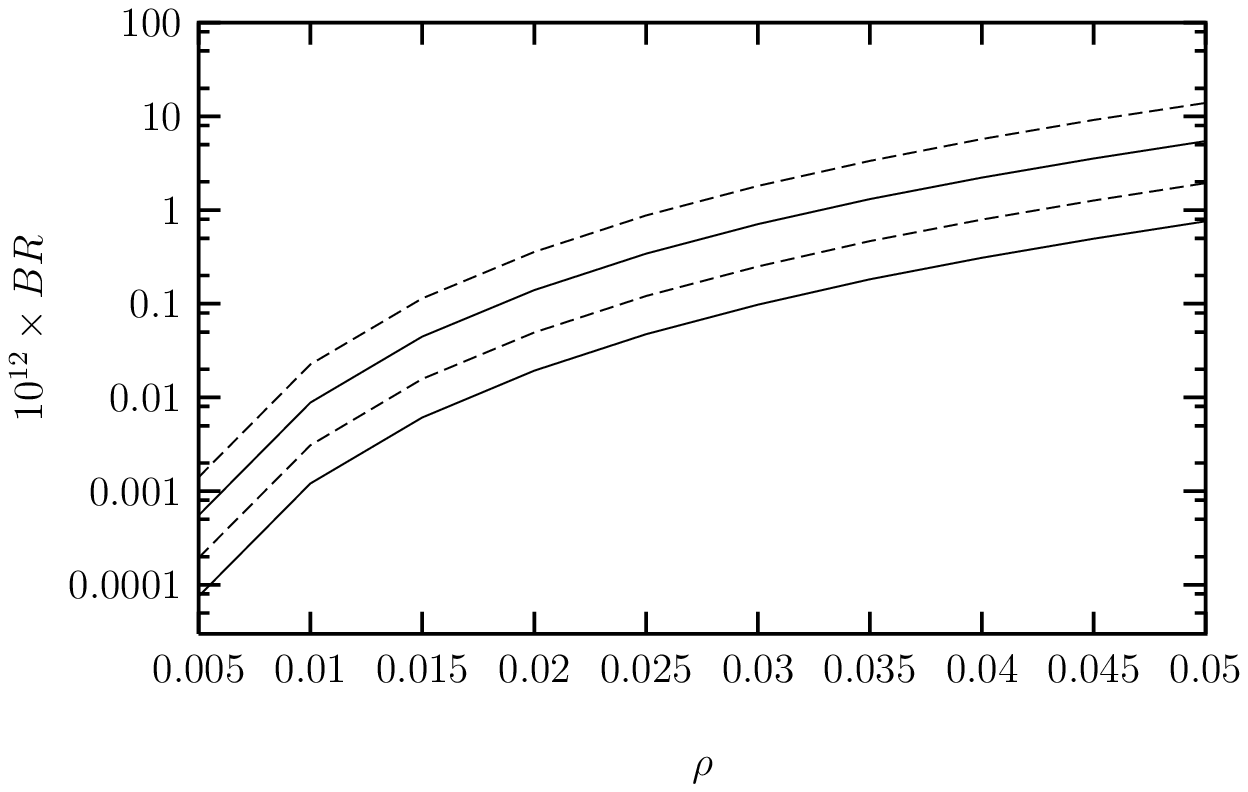} \vskip -3.0truein
\caption[]{The BR($\tau\rightarrow \mu \gamma$) with respect to
$\rho$. Here the lower-upper solid (dashed), line represents the
BR for a single extra dimension without-with lepton KK modes, for
$R=0.005\,GeV^{-1}$, the real couplings $\bar{\xi}^{E}_{N,\tau
\tau} =100\, GeV$, $\bar{\xi}^{E}_{N,\tau \mu} =50\, GeV$
($\bar{\xi}^{E}_{N,\tau \mu} =80\, GeV$).} \label{taumugamro}
\end{figure}

In Fig. \ref{tauegamro} and \ref{taumugamro} we present the
BR($\tau\rightarrow e \gamma$) and BR($\tau\rightarrow \mu
\gamma$) with respect to $\rho$, for different values of the
Yukawa couplings. Here the lower-upper solid (dashed), line
represents the BR for a single extra dimension without-with lepton
KK modes, for $R=0.005\,GeV^{-1}$, the real couplings
$\bar{\xi}^{E}_{N,\tau \tau} =100\, GeV$, $\bar{\xi}^{E}_{N,\tau
e} =0.1\, GeV$ ($\bar{\xi}^{E}_{N,\tau e} =0.5\, GeV$) and
$\bar{\xi}^{E}_{N,\tau \tau} =100\, GeV$, $\bar{\xi}^{E}_{N,\tau
\mu} =50\, GeV$ ($\bar{\xi}^{E}_{N,\tau \mu} =80\, GeV$). It is
observed that the BR is sensitive to the Gaussian width of the
localized neutral Higgs scalars and there is an enhancement at the
order of $10^4$ in the range $0.005\leq \rho \leq 0.05$ similar to
the previous decay. The BR is at the order of the magnitude of
$10^{-18}$ ($10^{-16}$) and $10^{-13}$ ($10^{-12}$) for the
intermediate values of the localization parameter $\rho$ and the
coupling $\bar{\xi}^{E}_{N,\tau e} =0.1\, GeV$
($\bar{\xi}^{E}_{N,\tau e} =0.5\, GeV$) and $\bar{\xi}^{E}_{N,\tau
\mu} =0.1\, GeV$ ($\bar{\xi}^{E}_{N,\tau \mu} =0.5\, GeV$). Notice
that the inclusion of lepton KK modes brings additional
enhancement at one order, in both decays.

\begin{figure}[htb]
\vskip -3.0truein \centering \epsfxsize=6.8in
\leavevmode\epsffile{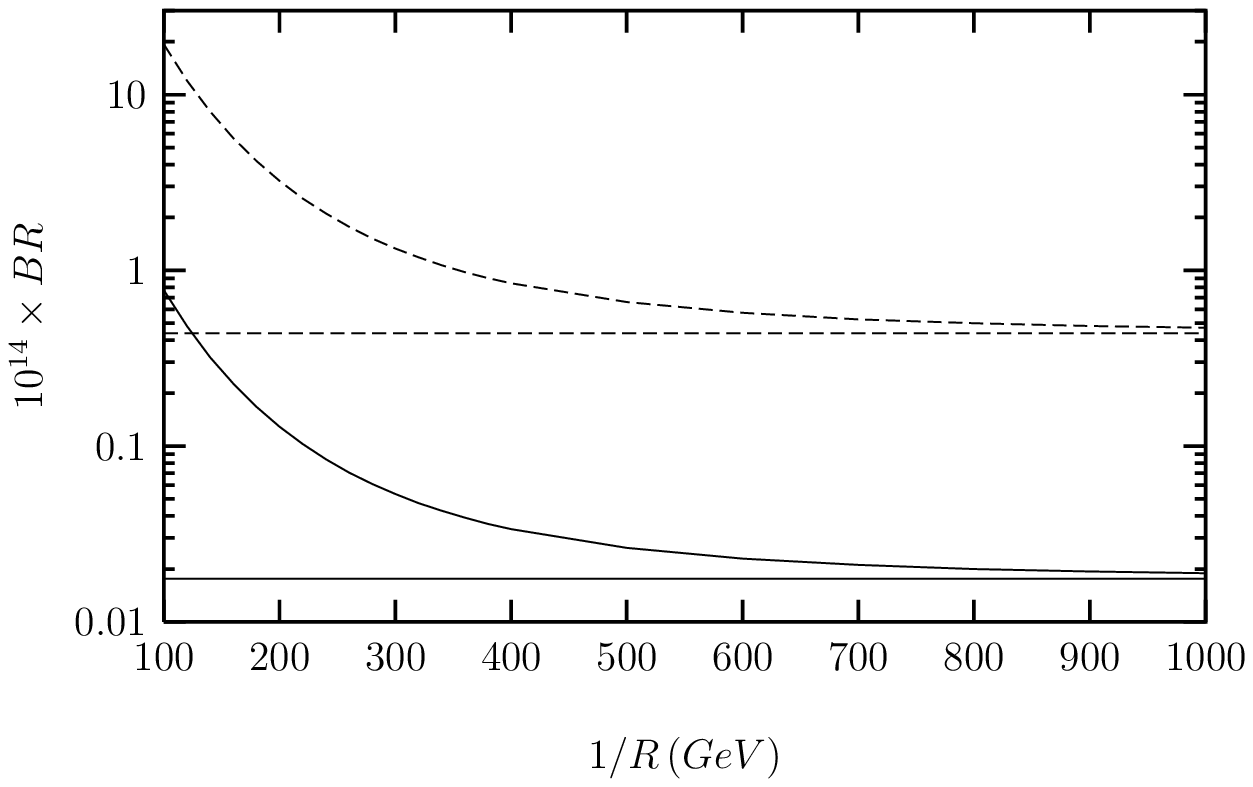} \vskip -3.0truein \caption[]{The
BR($\mu\rightarrow e \gamma$) with respect to the compactification
scale $1/R$. Here the lower-upper solid (dashed), line represents
the BR for a single extra dimension without-with lepton KK modes,
the real couplings $\bar{\xi}^{E}_{N,\tau \mu} =50\, GeV$,
$\bar{\xi}^{E}_{N,\tau e} =0.1\, GeV$ ($\bar{\xi}^{E}_{N,\tau e}
=0.5\, GeV$) and $\rho=0.01$. } \label{muegamR}
\end{figure}

\begin{figure}[htb]
\vskip -3.0truein \centering \epsfxsize=6.8in
\leavevmode\epsffile{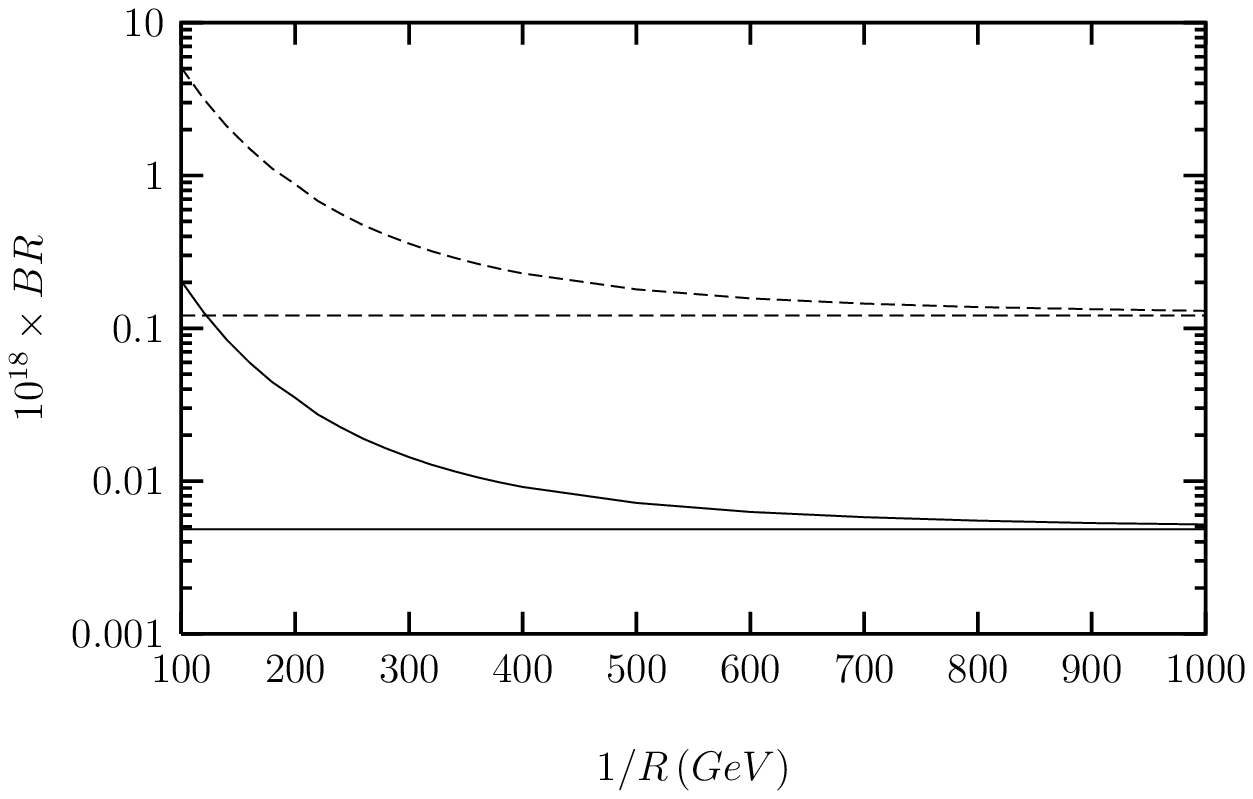} \vskip -3.0truein
\caption[]{BR($\tau\rightarrow e \gamma$) with respect to the
compactification scale $1/R$. Here the lower-upper solid (dashed),
line represents the BR for a single extra dimension without-with
lepton KK modes, the real couplings $\bar{\xi}^{E}_{N,\tau \tau}
=100\, GeV$, $\bar{\xi}^{E}_{N,\tau e} =0.1\, GeV$
($\bar{\xi}^{E}_{N,\tau e} =0.5\, GeV$) and $\rho=0.01$.  }
\label{tauegamR}
\end{figure}

\begin{figure}[htb]
\vskip -3.0truein \centering \epsfxsize=6.8in
\leavevmode\epsffile{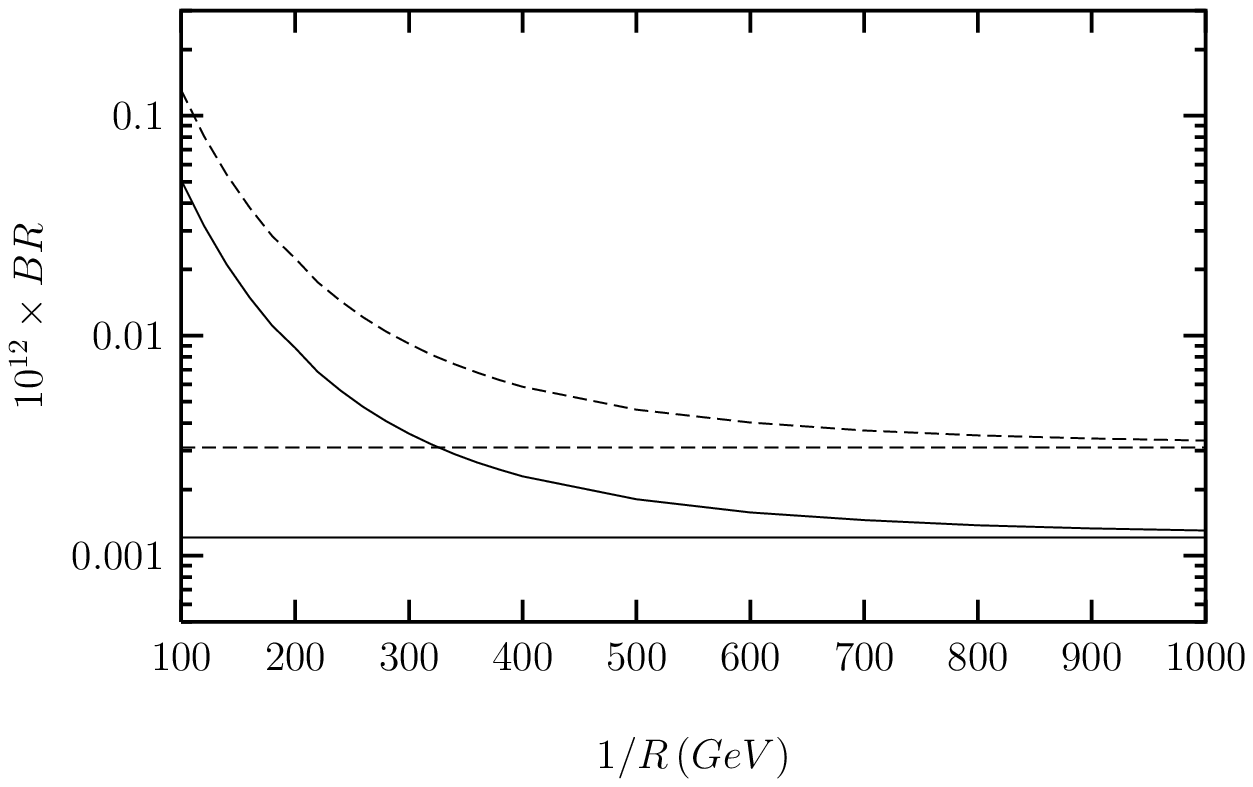} \vskip -3.0truein \caption[]{
BR($\tau\rightarrow \mu \gamma$) with respect to the
compactification scale $1/R$. Here the lower-upper solid (dashed),
line represents the BR for a single extra dimension without-with
lepton KK modes, the real couplings $\bar{\xi}^{E}_{N,\tau \tau}
=100\, GeV$, $\bar{\xi}^{E}_{N,\tau \mu} =50\, GeV$
($\bar{\xi}^{E}_{N,\tau \mu} =80\, GeV$) and $\rho=0.01$. }
\label{taumugamR}
\end{figure}

Figs. \ref{muegamR}; \ref{tauegamR}; \ref{taumugamR} are devoted
to the BR($\mu\rightarrow e \gamma$); BR($\tau\rightarrow e
\gamma$); BR($\tau\rightarrow \mu \gamma$) with respect to the
compactification scale $1/R$. Here the lower-upper solid (dashed),
line represents the BR for a single extra dimension without-with
lepton KK modes, the real couplings $\bar{\xi}^{E}_{N,\tau \mu}
=50\, GeV$, $\bar{\xi}^{E}_{N,\tau e} =0.1\, GeV$
($\bar{\xi}^{E}_{N,\tau e} =0.5\, GeV$); $\bar{\xi}^{E}_{N,\tau
\tau} =100\, GeV$, $\bar{\xi}^{E}_{N,\tau e} =0.1\, GeV$
($\bar{\xi}^{E}_{N,\tau e} =0.5\, GeV$); $\bar{\xi}^{E}_{N,\tau
\tau} =100\, GeV$, $\bar{\xi}^{E}_{N,\tau \mu} =50\, GeV$
($\bar{\xi}^{E}_{N,\tau \mu} =80\, GeV$) and $\rho=0.01$. These
figures show that the enhancement of the BR with the addition of
lepton KK modes is not so much sensitive to the compactification
scale $1/R$ for its large values.

Now, we study the effects of the position of the localization
point of the new Higgs doublet on the BR of the considered decays.

\begin{figure}[htb]
\vskip -3.0truein \centering \epsfxsize=6.8in
\leavevmode\epsffile{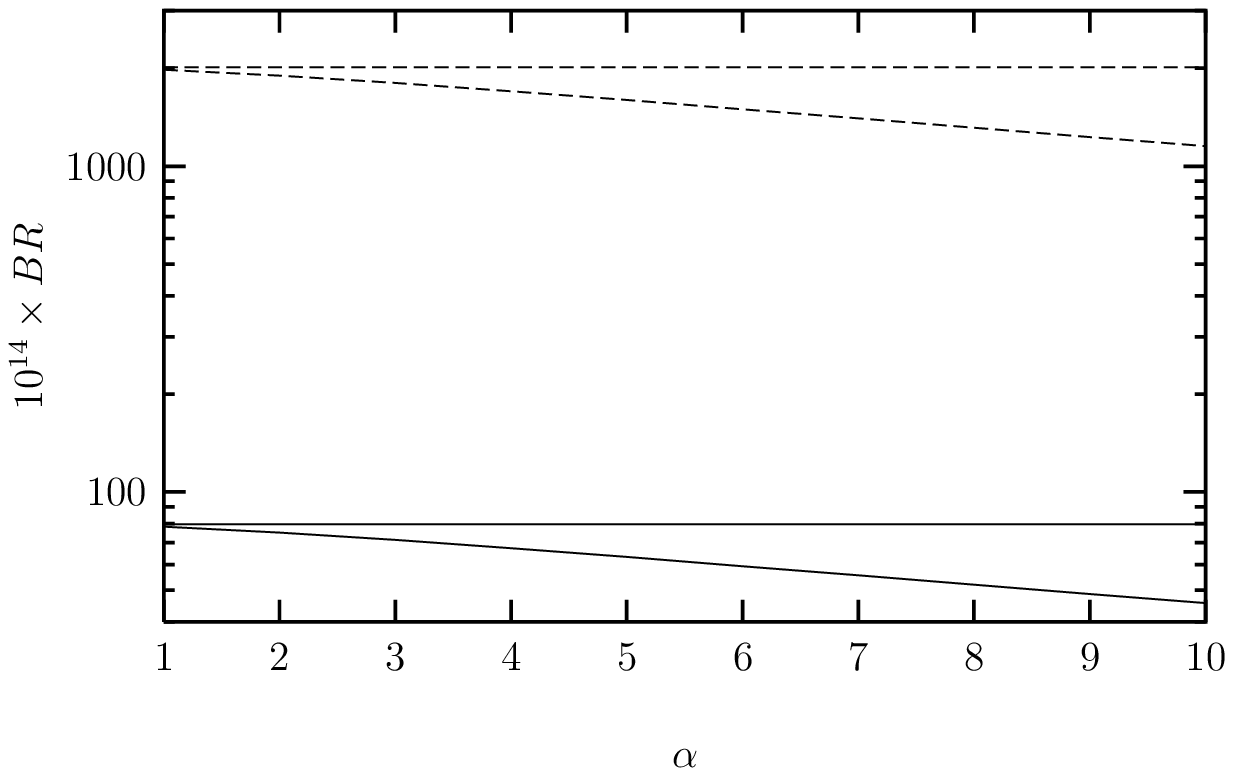} \vskip -3.0truein
\caption[]{The BR($\mu\rightarrow e \gamma$) with respect to
$\alpha$. Here the lower-upper solid (dashed), line represents the
BR for a single extra dimension with lepton KK modes,  for
$y_H=0-y_H=\alpha\, \sigma$, the real couplings
$\bar{\xi}^{E}_{N,\tau \mu} =50\, GeV$, $\bar{\xi}^{E}_{N,\tau e}
=0.1\, GeV$ ($\bar{\xi}^{E}_{N,\tau e} =0.5\, GeV$) and
$R=0.005\,GeV^{-1}$} \label{muegamroi}
\end{figure}

Fig. \ref{muegamroi} represents the BR($\mu\rightarrow e \gamma$)
with respect to $\alpha$, for different values of the Yukawa
couplings $\bar{\xi}^{E}_{N,\tau \mu}$ and $\bar{\xi}^{E}_{N,\tau
e}$. Here the lower-upper solid (dashed), line represents the BR
for a single extra dimension with lepton KK modes,  for
$y_H=0-y_H=\alpha\, \sigma$, the real couplings
$\bar{\xi}^{E}_{N,\tau \mu} =50\, GeV$, $\bar{\xi}^{E}_{N,\tau e}
=0.1\, GeV$ ($\bar{\xi}^{E}_{N,\tau e} =0.5\, GeV$) and
$R=0.005\,GeV^{-1}$. This figure shows that the  BR decreases with
the increasing values of $\alpha$, at the order of $\% 45$ in the
interval $1\leq \alpha \leq 10$, for the large values of the
Yukawa coupling.

\begin{figure}[htb]
\vskip -3.0truein \centering \epsfxsize=6.8in
\leavevmode\epsffile{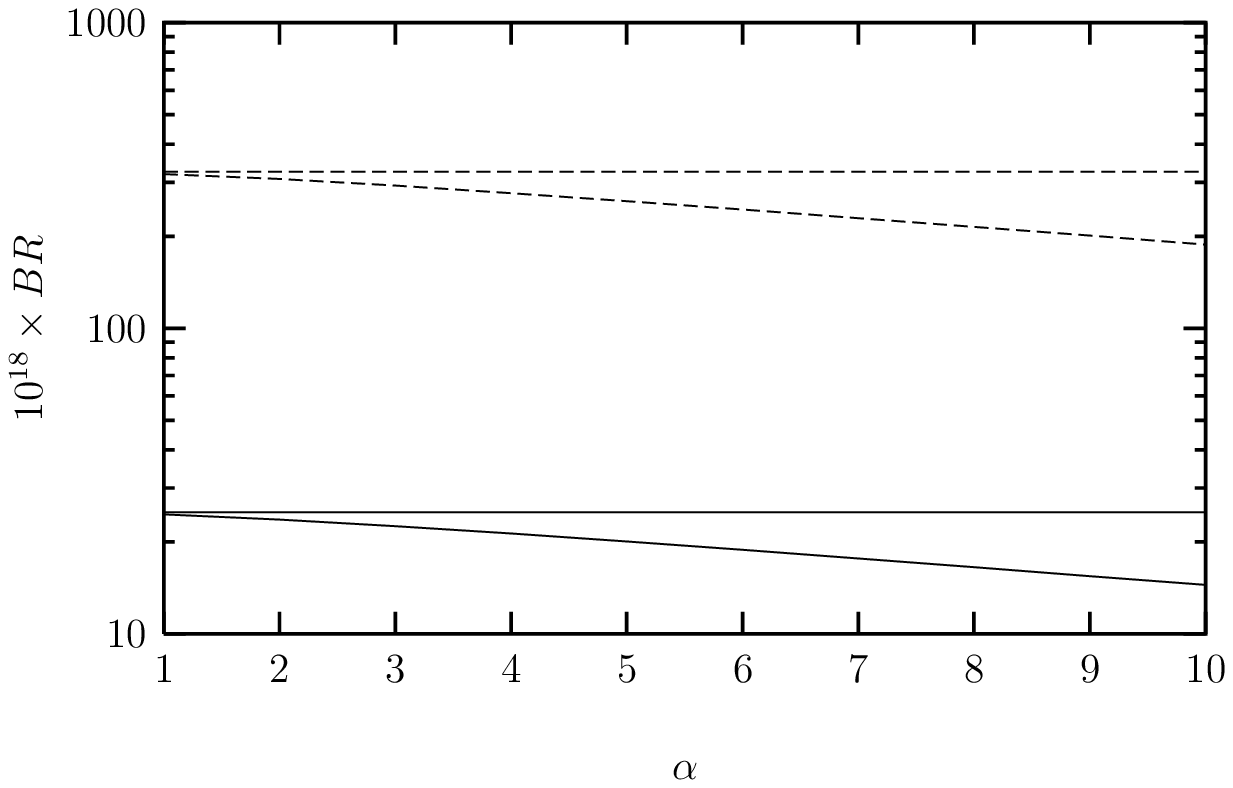} \vskip -3.0truein \caption[]{
The BR($\tau\rightarrow e \gamma$) with respect to $\alpha$. Here
the lower-upper solid (dashed), line represents the BR for a
single extra dimension with lepton KK modes,  for
$y_H=0-y_H=\alpha\, \sigma$, the real couplings
$\bar{\xi}^{E}_{N,\tau \tau} =100\, GeV$, $\bar{\xi}^{E}_{N,\tau
e} =0.1\, GeV$ ($\bar{\xi}^{E}_{N,\tau e} =0.5\, GeV$) and
$R=0.005\,GeV^{-1}$.} \label{tauegamroi}
\end{figure}

\begin{figure}[htb]
\vskip -3.0truein \centering \epsfxsize=6.8in
\leavevmode\epsffile{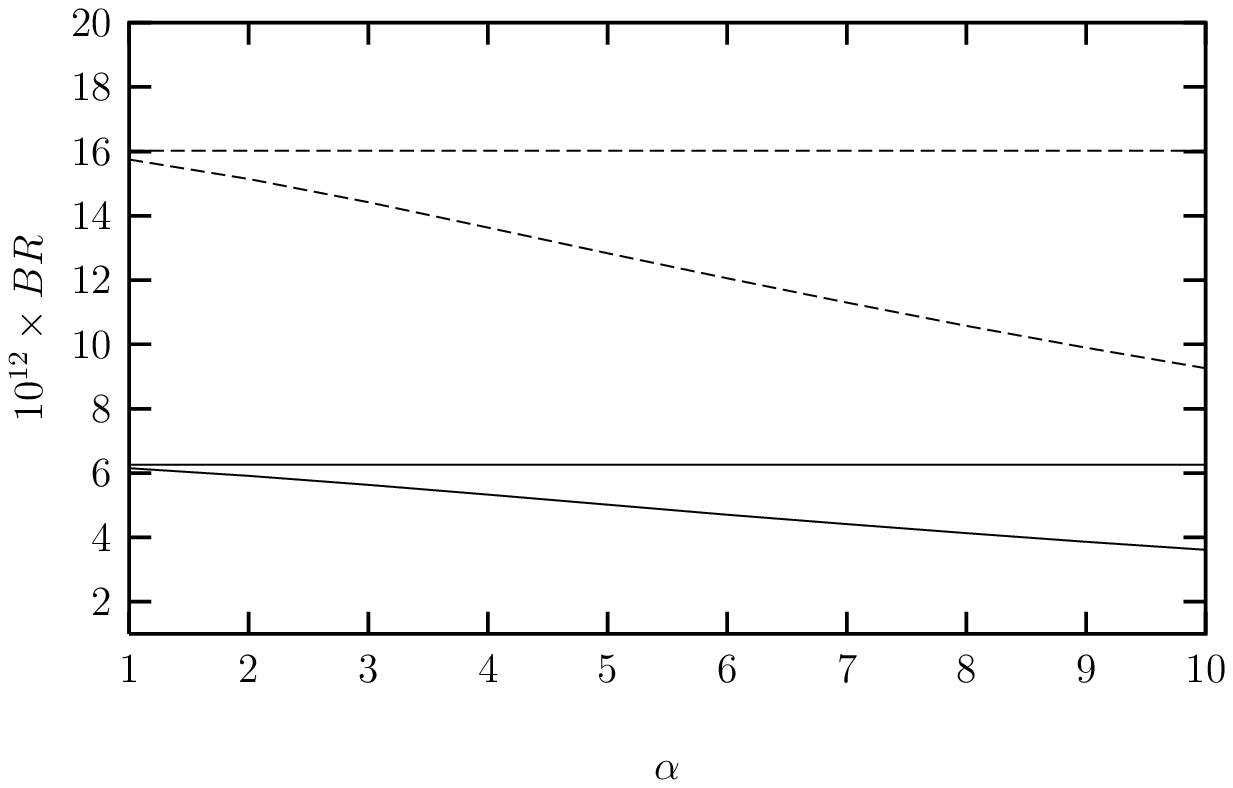} \vskip -3.0truein \caption[]{
The BR($\tau\rightarrow \mu \gamma$) with respect to $\alpha$.
Here the lower-upper solid (dashed), line represents the BR for a
single extra dimension with lepton KK modes,  for
$y_H=0-y_H=\alpha\, \sigma$, the real couplings
$\bar{\xi}^{E}_{N,\tau \mu} =50\, GeV$ ($\bar{\xi}^{E}_{N,\tau
\mu} =80\, GeV$) and $R=0.005\,GeV^{-1}$.} \label{taumugamroi}
\end{figure}

In Fig. \ref{tauegamroi}; \ref{taumugamroi} we present the
BR($\tau\rightarrow e \gamma$); BR($\tau\rightarrow \mu \gamma$)
with respect to $\alpha$. Here the lower-upper solid (dashed),
line represents the BR for a single extra dimension with lepton KK
modes,  for $y_H=0-y_H=\alpha\, \sigma$, the real couplings
$\bar{\xi}^{E}_{N,\tau \tau} =100\, GeV$, $\bar{\xi}^{E}_{N,\tau
e} =0.1\, GeV$ ($\bar{\xi}^{E}_{N,\tau e} =0.5\, GeV$);
$\bar{\xi}^{E}_{N,\tau \tau} =100\, GeV$, $\bar{\xi}^{E}_{N,\tau
\mu} =50\, GeV$ ($\bar{\xi}^{E}_{N,\tau \mu} =80\, GeV$) and
$R=0.005\,GeV^{-1}$. These figures show that the BR decreases with
the increasing values of $\alpha$, at the order of $\% 35$; $\%
45$ in the interval $1\leq \alpha \leq 10$, for the large values
of the Yukawa coupling.

At this stage we would like to summarize our results:
\begin{itemize}
\item  The BR is strongly sensitive to the Gaussian width of the
localized neutral Higgs scalars and it increases with the
increasing values of the width for the decays under consideration.
This enhancement is almost at the order of $10^4$ in the range
$0.005\leq \rho \leq 0.05$. The inclusion of lepton KK modes
brings additional enhancement at one order. The BR for
$\mu\rightarrow e \gamma$ ($\tau\rightarrow e \gamma$,
$\tau\rightarrow \mu \gamma$) is at most at the order of the
magnitude of $10^{-12}$ ($10^{-16}$, $10^{-12}$) for the
intermediate values of the localization parameter $\rho$ and the
couplings taken.
\item The BR decreases with the increasing distance,
$y_H=\alpha\,\sigma$, of the localization point of the new Higgs
doublet from the origin in the extra dimension. With the
increasing values of $\alpha$, there is almost $\% 50$ suppression
of the BRs of the decays under consideration in the interval
$1\leq \alpha \leq 10$, for the large values of the Yukawa
couplings.
\end{itemize}
The improvement of the experimental results of the radiative LFV
decay BRs would make it possible to search the effects of the
extra dimension and the possible localization of the new Higgs
bosons in the extra dimension.
\section{Acknowledgement}
I thank my father Oguz Iltan for his spiritual support.
\newpage
\newpage

\begin{thebibliography}{1}
%
\bibitem{Brooks} M. L. Brooks et. al., MEGA Collaboration,
{\it Phys. Rev. Lett.} {\bf 83}, 1521 (1999).
%
\bibitem{Hayasaka} K. Hayasaka et al.., {\it Phys. Lett.} {\bf B613}
(2005) 20.
%
\bibitem{Nicolo} Donato Nicolo, MUEGAMMA Collaboration,
{\it Nucl. Instrum. Meth} {\bf A503} (2003) 287.
%
\bibitem{Yamada} S. Yamada, {\it Nucl. Phys. Proc. Suppl.} {\bf 144} (2005)
185.
%
%
\bibitem{Roney} J.M. Roney and the BABAR Collaboration,
{\it Nucl. Phys. Proc. Suppl.} {\bf 144} (2005) 155.
%
\bibitem{Aubert} B. Aubert et. al., BABAR Collaboration,
SLAC-PUB-11028, BABAR-PUB-04-049, Feb. 2005, 7. pp, {\it Phys.
Rev. Lett.} {\bf 95} (2005) 041802.
%
\bibitem{Barbieri1} R. Barbieri and L. J. Hall,
{\it Phys. Lett.} {\bf B338}, 212 (1994); R. Barbieri, L. J. Hall
and A. Strumia, {\it Nucl. Phys.} {\bf B445}, 219 (1995);
 R. Barbieri, L. J. Hall and A. Strumia,
{\it Nucl. Phys.} {\bf B449}, 437 (1995); P. Ciafaloni, A.
Romanino and A. Strumia, IFUP-YH-42-95; T. V. Duong, B. Dutta and
E. Keith, {\it Phys. Lett.} {\bf B378}, 128 (1996); G. Couture,
et. al., {\it Eur. Phys. J.} {\bf C7}, 135 (1999); Y. Okada, K.
Okumara and Y. Shimizu, {\it Phys. Rev.} {\bf D61}, 094001 (2000).
%
\bibitem{Iltan1} E. O. Iltan, {\it Phys. Rev.} {\bf D64}, 115005, (2001);
{\it Phys. Rev.} {\bf D64} 013013, (2001).
%
\bibitem{Diaz} R. Diaz, R. Martinez and J-Alexis Rodriguez,
Phys.Rev. D63 (2001) 095007.
%
\bibitem{IltanExtrDim} E. O. Iltan, {\it JHEP} {\bf 0408},
20, (2004).
%
\bibitem{IltanLFVSplit} E. O. Iltan, hep-ph/0504013, (2005).
%
\bibitem{IltanLFVSplitFat} E. O. Iltan, hep-ph/0509096, (2005).
%
\bibitem{Chang} D. Chang, W. S. Hou and W. Y. Keung,
{\it Phys. Rev.} {\bf D48}, 217 (1993).
%
\bibitem{Paradisi} P. Paradisi, hep-ph/0508054, (2005).
%
%
\bibitem{Arkani} N. Arkani-Hamed, S. Dimopoulos, and G. R. Dvali,
{\it Phys. Rev.} {\bf D59} 086004 (1999); N. Arkani-Hamed, S.
Dimopoulos, and G. R. Dvali, {\it Phys. Lett.} {\bf B429}, 263
(1998);  .
%
\bibitem{Antoniadis1} I. Antoniadis,{\it Phys. Lett.} {\bf B246},
377 (1990); I.Antoniadis and K. Benakli, {\it Phys. Lett.} {\bf
B326}, 69 (1994).
%
\bibitem{Antoniadis2}
I. Antoniadis, K. Benakli, and M. Quiros, {\it Phys. Lett.} {\bf
B331}, 313 (1994); A. Pomarol and M. Quiros, {\it Phys. Lett.}
{\bf B438}, 255 (1998); I. Antoniadis, K. Benakli, and M. Quiros,
{\it Phys. Lett.} {\bf B460}, 176 (1999); P. Nath, Y. Yamada, and
M. Yamaguchi, {\it Phys. Lett.} {\bf B466}, 100 (1999); M. Masip
and A. Pomarol, {\it Phys. Rev.} {\bf D60}, 096005 (1999) A.
Delgado, A. Pomarol, and M. Quiros, {\it JHEP} {\bf 01}, 030
(2000); T. G. Rizzo and J. D. Wells, {\it Phys. Rev.} {\bf D61},
016007 (2000); P. Nath and M. Yamaguchi, {\it Phys. Rev.} {\bf
D60}, 116004 (1999); A. Muck, A. Pilaftsis, and R. Ruckl, {\it
Phys. Rev.} {\bf D65}, 085037 (2002); A. Muck, A. Pilaftsis, and
R. Ruckl, hep-ph/0203032; C. D. Carone, {\it Phys. Rev.} {\bf
D61}, 015008 (2000).

\bibitem{Antoniadis3}I. Antoniadis, C. Munoz, M. Quiros,
{\it Nucl. Phys} {\bf B397} 515 (1993);
%
\bibitem{Appelquist} T. Appelquist, H.-C. Cheng, B. A. Dobrescu,
 {\it Phys. Rev.} {\bf D64} 035002 (2001).
%
\bibitem{Papavassiliou}
J. Papavassiliou and A. Santamaria, {\it Phys. Rev.} {\bf D63},
016002 (2001).
%
\bibitem{Chakraverty}
D. Chakraverty, K. Huitu, and A. Kundu, Phys.Lett. B558 (2003)
173-181; A. J. Buras, M. Spranger, and A. Weiler, Nucl.Phys. B660
(2003) 225.
%
\bibitem{Agashe} K. Agashe, N.G. Deshpande, G.-H. Wu, {\it Phys. Lett.}
{\bf 514} 309 (2001).
%
\bibitem{Dienes} K. R. Dienes, E. Dudas, T. Gherghetta
9811428, Q. H. Shrihari, S. Gopalakrishna, C. P. Yuan, 0312339
%
\bibitem{Agulia} F. Agulia, M. P. Victoria, J. Santiago,
{\it JHEP} {\bf 0302} 051, (2003); {\it Acta Phys.Polon.} {\bf
B34}, 5511 (2003).
%
\bibitem{iltanEDM} E. O. Iltan, hep-ph/0401229; {\it JHEP} {\bf
0402} 065, (2004); {\it JHEP} {\bf 0408} 020, (2004); {\it JHEP}
{\bf 0404} 018, (2004); {\it Mod. Phys. Lett.} {\bf A20} 1845,
(2005); {\it Eur. Phys. J.} {\bf C41} 233, (2005).
%
\bibitem{Lam} C. S. Lam, hep-ph/0302227, 2003; C. A. Scrucca, M. Serona,
L. Silvestrini, {\it Nucl.Phys.} {\bf B669} 128, (2003); M. Gozdz,
W. A. Kaminsk, {\it Phys. Rev.} {\bf D68} 057901, (2003) ; C.
Biggio, et.al, {\it Nucl.Phys.} {\bf B677} 451, (2004); M. Carena,
et.al, {\it Phys. Rev.} {\bf D68} 035010, (2003) ; A. J. Buras,
et. al., {\it Nucl.Phys.} {\bf B678} (2004) 455; T. G. Rizzo, {\it
JHEP} {\bf 0308} 051 (2003), A. J. Buras, hep-ph/0307202, 2003; S.
Matsuda, S. Seki, hep-ph/0307361, 2003; R. N. Mohapatra, {\it
Phys. Rev.} {\bf D68} 116001, (2003); B. Lillie, {\it JHEP} {\bf
0312} 030 (2003); A.A Arkhipov, hep-ph/0309327 2003; F.Feruglia,
{\it Eur.Phys.J.} {\bf C33} (2004) S114.
%
\bibitem{Mirabelli}
E. A. Mirabelli, Schmaltz, {\it Phys. Rev.} {\bf D61} (2000)
113011.
%
\bibitem{Changg} W. F. Chang, I. L. Ho and J. N. Ng, {\it Phys. Rev.}
{\bf D66} 076004 (2002).
%
\bibitem{Branco}
G. C. Branco,A. Gouvea, M. N. Rebelo,  {\it Phys. Lett.} {\bf
B506} (2001) 115.
%
\bibitem{Chang2}
W. F. Cang, J. N. Ng, {\it JHEP} {\bf 0212} (2002) 077.
%
\bibitem{Hewett} B. Lillie, J. L. Hewett, {\it Phys. Rev.} {\bf 68}
(2003) 116002.
%
\bibitem{Perez}
Y. Grossman and G. Perez, {\it Phys. Rev.} {\bf D67} (2003)
015011; {\it Pramana} {\bf 62} (2004) 733.
%
\bibitem{IltanEDMSplit} E. O. Iltan, {\it Eur.Phys.J.} {\bf C44} (2005)
411.
%
\bibitem{Grossman}
Y. Grossman, {\it Int. J. Mod. Phys.} {\bf A15} (2000) 2419; D. E.
Kaplan and T. M. Tait, {\it JHEP} {\bf 0111} (2001) 051; G.
Barenboim,  {\it et. al.}, {\it Phys. Rev.} {\bf D64} (2001)
073005; W. F. Chang, I. L. Ho and J. N. Ng, {\it Phys. Rev.} {\bf
D66} (2002) 076004; W. Skiba and D. Smith, {\it Phys. Rev.} {\bf
D65} (2002) 095002; Y. Grossman, R. Harnik, G. Perez, M. D.
Schwartz and Z. Surujon, {\it Phys. Rev.} {\bf D71} (2005) 056007;
P. Dey, G. Bhattacharya, {\it Phys. Rev.} {\bf D70} (2004) 116012;
Y. Nagatani, G. Perez, {\it JHEP} {\bf 0502} (2005) 068; R.
Harnik, G. Perez, M. D. Schwartz, Y. Shirman, {\it JHEP} {\bf
0503} (2005) 068.
%
\bibitem{Surujon} Z. Surujon, {\it Phys.Rev.} {\bf D73} (2006) 016008.
%
\bibitem{iltSplitHiggsLocal} E. O. Iltan, hep-ph/0511241
%
\bibitem{Sher} T. P. Cheng and M. Sher, {\it Phy. Rev.} {\bf D35} (1987) 3383.
%
\bibitem{Rizzo} T. G. Rizzo, J. D. Wells, {\it Phy. Rev.} {\bf D61}
(2000) 016007.
%
\end{thebibliography}
\end{document}